\documentclass[12pt]{article}

\usepackage{scicite}

\usepackage{amsmath}
\usepackage{amsfonts}
\usepackage{amssymb}
\usepackage{graphicx}
\usepackage{braket}
\usepackage{comment}
\usepackage{color}

\newcommand{\tr}[2][]{\text{Tr}_{#1}\left[#2\right]}
\newcommand{\probsU}[2][]{\ensuremath{{P}\left(#2\right)^{#1} }}

\newenvironment{sciabstract}{%
\begin{quote} \bf}
{\end{quote}}

\title{Probing entanglement entropy via randomized measurements}

\author
{Tiff Brydges,${}^{1,2\ast}$, Andreas Elben,${}^{1,3\ast}$, Petar Jurcevic,${}^{1,2}$,\\ Beno\^{i}t Vermersch,${}^{1,3}$, Christine Maier,${}^{1,2}$, Ben P. Lanyon,${}^{1,2}$,\\ Peter Zoller,${}^{1,3}$, Rainer Blatt,${}^{1,2}$, Christian F. Roos  ${}^{1,2\dagger}$\\
\\
\normalsize{${}^{1}$Institute for Quantum Optics and Quantum Information of the Austrian Academy of Sciences,}\\
\normalsize{A-6020 Innsbruck, Austria}\\
\normalsize{${}^{2}$Institute for Experimental Physics, University of Innsbruck, A-6020 Innsbruck, Austria}\\
\normalsize{${}^{3}$Institute for Theoretical Physics, University of Innsbruck, A-6020 Innsbruck, Austria}\\
\\
\normalsize{$^\ast$ These authors contributed equally.}
\\
\normalsize{$^\dagger$To whom correspondence should be addressed; E-mail:  christian.roos@uibk.ac.at.}
}

\date{}

\begin{document} 




\maketitle 

\begin{sciabstract}
Entanglement is the key feature of many-body quantum systems, and the development of new tools to probe it in the laboratory is an outstanding challenge. 
Measuring the entropy of different partitions of a quantum system provides a way to probe its entanglement structure. 
Here, we present and experimentally demonstrate a new protocol for measuring entropy, based on statistical correlations between randomized measurements. 
Our experiments, carried out with a trapped-ion quantum simulator, prove the overall coherent character of the system dynamics and reveal the growth of entanglement between its parts - both in the absence and presence of disorder. 
Our protocol represents a universal tool for probing and characterizing engineered quantum systems in the laboratory, applicable to arbitrary quantum states of up to several tens of qubits.
\end{sciabstract}


Engineered quantum systems, consisting of tens of individually-controllable interacting quantum particles, are currently being developed using a number of different physical platforms; including atoms in optical arrays
\cite{Bloch:2012,Browaeys:2016,Saffman:2016},
 ions in radio-frequency traps \cite{Zhang:2017a,Friis:2018}, and superconducting circuits \cite{Fitzpatrick:2017,Gambetta:2017,Xu:2018,Neill:2018}.
These systems offer the possibility of generating and probing complex quantum states and dynamics particle by particle - finding application in the near-term as quantum simulators, and in the longer-term as quantum computers. As these systems are developed, new protocols are required to characterize them - to verify that they are performing as desired and to measure quantum phenomena of interest. 

A key property to measure in engineered quantum systems is entanglement. For example, in order for quantum simulators and computers to provide an advantage over their classical analogues, they must generate large amounts of entanglement between their parts \cite{Vidal:2003}. Furthermore, when using these devices to tackle open questions in physics, the dynamics of entanglement provides signatures of the phenomena of interest, such as thermalization~\cite{Calabrese:2005} and many-body localization~\cite{Basko:2006,Abanin:2018}.

Entanglement can be probed by measuring entanglement entropies. In particular, consider the second-order R\'enyi entropy
\begin{equation}
S^{(2)}(\rho_A)=-\log_2\mbox{Tr}(\rho_A^2),\
\label{eq:Renyi}
\end{equation}
with $\rho_A$ the reduced density matrix for a part $A$ of the total system described by $\rho$. 
If the entropy of part $A$ is greater than the entropy of the total system; i.e $S^{(2)}(\rho_A)>S^{(2)}(\rho)$, bipartite entanglement exists between $A$ and the rest of the system \cite{Horodecki:2009}.
Thus, a measurement of the entropy of the whole system, as well as of its subsystems, provides information about the entanglement contained within the system. Additionally, a measurement of the entropy of the total state $\rho$ provides the opportunity  to verify the overall coherence of the system, as for pure quantum states $S^{(2)}(\rho)=0$.

Recently, a protocol to directly measure the second-order R\'enyi entropy, $S^{(2)}$, has been demonstrated, requiring collective measurements to be made on two identical copies $\rho$ of a quantum system \cite{Ekert:2002,Islam:2015, Kaufman:2016, Linke:2017}. In \cite{Kaufman:2016}, that protocol was used to study entanglement growth and thermalization in a six-site Bose-Hubbard system, realized with atoms in an optical lattice. 

In this work, we present and experimentally demonstrate a new protocol to measure the second-order R\'enyi entropy, $S^{(2)}$, based on, and extending, the proposals of \cite{vanEnk:2012, Elben:2018,Vermersch:2018,Elben:2018b}. 
Key strengths of the protocol are that it 
requires preparation of only a single copy of the quantum system at a time and can be implemented on any physical platform with single-particle readout and control.
While recently efficient tomographic methods \cite{Lanyon:2017, Torlai:2018} have been developed to characterize weakly entangled states, in contrast our approach imposes no a-priori assumption on the structure of the quantum state. 
In our experiments, we use the protocol to measure the dynamical evolution of entanglement entropy of up to 10-qubit partitions of a trapped-ion quantum simulator: a system size and values of measured entropy that have been beyond reach using other methods. 

The key insight of the protocol is that information about the second-order R\'enyi entropies of a system is contained in statistical correlations between the outcomes of measurements perfomed in random bases. Specifically, for a system of $N$ qubits, the approach~\cite{Elben:2018} is to apply a product of single-qubit unitaries $U=u_1 \otimes .. \otimes u_N$, where each unitary $u_i$ is drawn independently from the circular unitary ensemble (CUE) \cite{Mezzadri:2007}, and then to measure the qubits in a fixed (logical) basis. For each $U$, repeated measurements are made to obtain statistics, and the entire process is repeated for many different randomly drawn instances of $U$. The second-order R\'enyi entropy, $S^{(2)}$, of the density matrix $\rho_A$ for an arbitrary partition $A=\{i[1],..,i[N_A]\}$ of $N_A\le N$ qubits is then obtained from 
\begin{equation}
    S^{(2)}(\rho_A) = -\log_2\overline{X}, \;\;\mbox{with}\; X= 2^{N_A} \sum_{s_A,s_A'}  (-2)^{-D[s_A,s_A']} P(s_A) P(s_A'), \label{eq:Renyi-from-random-meas}
\end{equation}
where $\overline{\vphantom{P} \dots}$ denotes the ensemble average of (cross-) correlations of excitation probabilities $P(s_A)=\bra{s_A}U_A\rho_A U_A^\dagger\ket{s_A}$;  $s_A$ are the logical basis states of partition $A$, $U_A=U|_A$ the restriction of $U$ to $A$, and $D[s_A,s_A']$ is the Hamming distance between $s_A$ and $s'_A$. Note that $\overline{X}$ is equal to the purity $\mathrm{Tr}(\rho_A^2)$ of the density matrix $\rho_A$.
We remark that Eq.~\eqref{eq:Renyi-from-random-meas} represents an explicit formula, proven in the supplement~\cite{SupplementaryMaterial}, to reconstruct the second-order R\'enyi entropy of the subsystem of interest directly. As a result, compared to the recursive scheme presented in Ref.~\cite{Elben:2018}, an exponential overhead in the classical postprocessing is avoided.

\begin{figure}
    \centering
    \includegraphics[width=0.9\columnwidth]{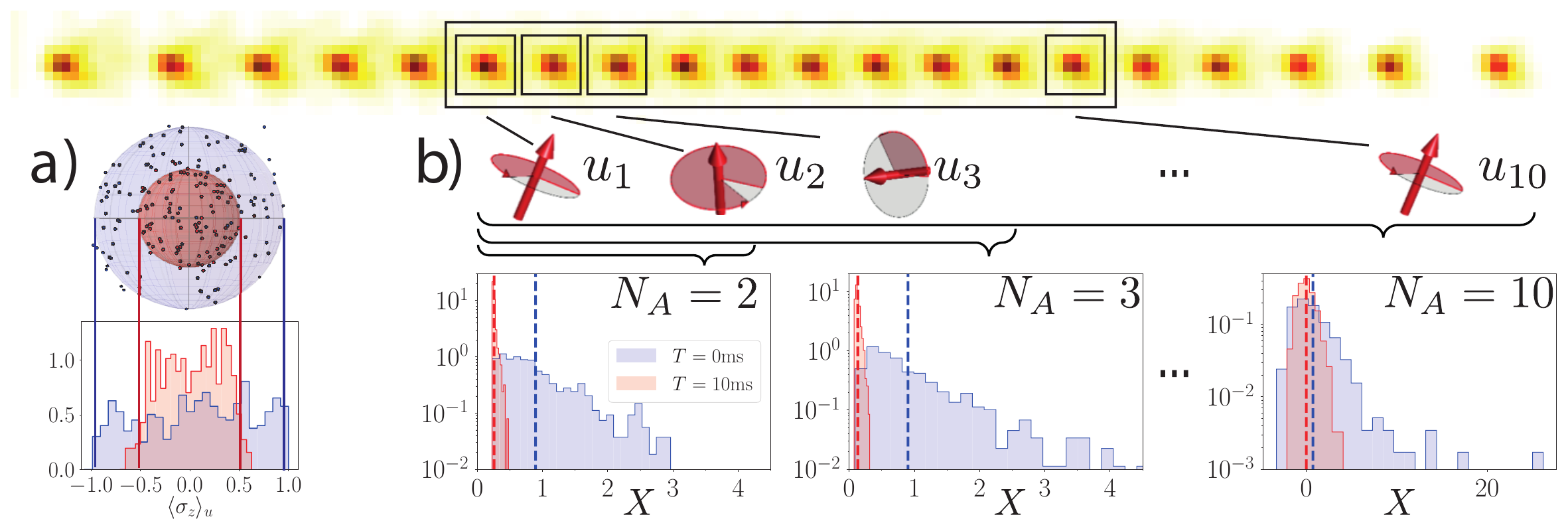}
    \caption{{\bf Measuring second-order R\'enyi entropies via randomized measurements.} a) Single qubit Bloch sphere. The purity is directly related to the width of the distribution of measurement outcomes after applying random rotations $u_i$. Initial pure state (blue) and mixed state (red)  cases are shown. See text. b) Generalization to multiple qubits: Measuring up to 10-qubit partitions of a 20-qubit string, as shown (top).
    Repeated measurements are made to obtain statistics, see text. Experimental data (bottom): Histograms of the weighted sum $X$ of cross correlations (as defined in Eq.~\eqref{eq:Renyi-from-random-meas}), with mean values corresponding to the purities (dashed lines). Results are shown for two different times during evolution under $H_{\mathrm{XY}}$, starting from a highly pure, separable state and evolving into a high entropy state.}
    \label{fig:1}
\end{figure}

For the partition of a single qubit, $N_A=1$, the Bloch sphere provides a simple graphical representation to understand the relation between the purities and the distribution of excitation probabilities (see Fig.~\ref{fig:1}a). For a pure state, $\mathrm{Tr}(\rho_A^2)=1$, the quantum state can be represented as a unit Bloch vector on the sphere, with random rotations leading to a uniform distribution of probabilities covering the full range $[0,1]$.  For a mixed state, $\mathrm{Tr}(\rho_A^2)<1$, the length of the Bloch vector is less than $1$, and the probabilities take values in a reduced interval. 
Generalizing to the multi-qubit scenario, the purities are directly inferred from the mean of the statistical distribution of a weighted sum of cross correlations using Eq.~\eqref{eq:Renyi-from-random-meas}.  Examples of cross correlations that were measured for different partition sizes of the trapped-ion system are shown in Fig.~\ref{fig:1}(b), together with the estimated purities.

Our experiments were implemented using strings of up to twenty trapped $^{40}$Ca$^+$ ions, each of which encodes a qubit that can be individually manipulated by spatially focused, coherent laser pulses. When dressed with suitably tailored laser fields, the ions are subject to a quantum evolution that is equivalent to a model of spins interacting via a long-range XY model \cite{Porras:2004} in the presence of a transverse field, 
\begin{equation}
    H_{\mathrm{XY}} = \hbar\sum_{i<j}J_{ij}(\sigma^{+}_{i}\sigma^{-}_{j}+\sigma^{-}_{i}\sigma^{+}_{j}) + \hbar B \sum_{j}\sigma^{z}_{j}\,.
    \label{eq:XY-Hamiltonian}
\end{equation}
Here, $\sigma_i^{\beta}$ ($\beta = x,y,z$) are the spin-$1/2$ Pauli operators, $\sigma_i^{+}(\sigma_i^{-})$ the spin-raising (lowering) operators acting on spin $i$, and $J_{ij} \approx J_0/ \lvert{i-j}\rvert^{\alpha}$ the coupling matrix with an approximate power-law decay and $0<\alpha<3$.
For further experimental details, see \cite{SupplementaryMaterial,Jurcevic:2014}. Optionally, a locally disordered potential could be added \cite{Smith:2016,Maier:2018}, realizing the Hamiltonian $H=H_{\mathrm{XY}}+H_\mathrm{D}$, with $H_{\mathrm{D}}=\hbar \sum_{j}\Delta_j\sigma^{z}_{j}$ and $\Delta_j$ the magnitude of disorder applied to ion $j$.
For entropy measurements, the following experimental protocol was used throughout: the system was initially prepared in the N\'eel ordered product state $\rho_{0}=|\psi\rangle\langle\psi|$ with $|\psi\rangle=|\!\downarrow\uparrow\downarrow..\!\uparrow\rangle$. This state was subsequently time-evolved under $H_{\mathrm{XY}}$ (or $H$) into the state $\rho(t)$.
The coherent interactions arising from this time evolution generated varying types of entanglement in the system. Subsequently, randomized measurements on $\rho(t)$ were performed through individual rotations of each qubit by a random unitary ($u_{i}$), sampled from the CUE~\cite{Mezzadri:2007}, followed by a state measurement in the $z$-basis. Each $u_{i}$ can be decomposed into three rotations $R_{z}(\theta_{3})R_{y}(\theta_{2})R_{z}(\theta_{1})$, and two random unitaries were concatenated to ensure that drawing of the $u_{i}$ was stable against small drifts of physical parameters controlling the rotation angles $\theta_i$ \cite{SupplementaryMaterial}.
Finally, spatially resolved fluorescence measurements realised a projective measurement in the logical z-basis. To measure the entropy of a quantum state, $N_{U}$ sets of single-qubit random unitaries, $U=u_1 \otimes \dots \otimes u_N$, were applied. For each set of applied unitaries, $U$, the measurement was repeated $N_{M}$ times.

\begin{figure}
    \centering
    \includegraphics[width=.7\columnwidth]{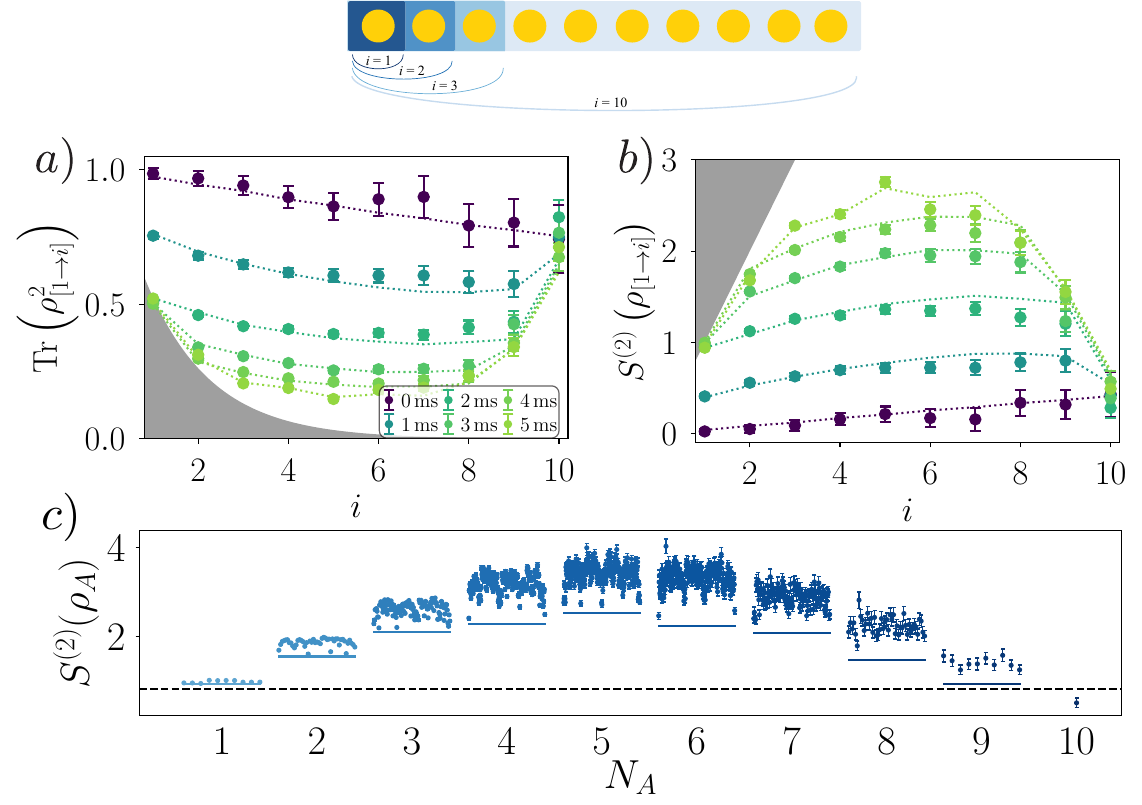}
    \caption{{\bf Purity and second-order R\'enyi entropies of a 10-qubit system}.  Measured purity, (a), and  second-order R\'enyi entropy, (b), of a N\'eel state, time-evolved under $H_{\mathrm{XY}}$ ($J_0=420\,$s$^{-1}$, $\alpha=1.24$), for connected partitions $[1\rightarrow i]$. Dotted curves are purities derived from a numerical simulation. Maximally mixed states with minimal purity fall on the boundary of the shaded area. (c) Second-order R\'enyi entropy, $S^{(2)}(\rho_A)$, of all $2^{10}-1=1023$ partitions at $t=5$ ms, with $N_A$ denoting the number of ions in a partition $A$.  For all data points, $N_{M}=150$ and $N_U=500$. Error bars, which increase with subsystem size \cite{SupplementaryMaterial}, are standard errors of the mean $\overline{X}$. Lines in (c) are drawn at three standard errors above the full system's entropy (black, dashed) and below the minimal subsystem's entropy (blue, solid).}
    \label{fig:10-qubit Purity-and-entropy vs partition size}
\end{figure}

In the first experiment, the 10-qubit state $\rho_0$ was prepared and subsequently time-evolved under $H_{\mathrm{XY}}$ (Eq.~\eqref{eq:XY-Hamiltonian}), without disorder, for $\tau=0,\ldots, 5$~ms. 
Fig.~\ref{fig:10-qubit Purity-and-entropy vs partition size} shows the measured purities (a) and entropies (b) of all connected partitions that include qubit $1$ during this quench. 
The overall purity (and thus entropy) remained at a constant value of $\tr[]{\rho^2}=0.74\pm 0.07$, within error, throughout the time evolution, implying that the time evolution was approximately unitary. The initial state's reconstructed purity is in agreement with control experiments, which show a purity loss of 0.08 due to imperfect state preparation and an underestimation of the purity by approximately 0.17 due to decoherence during the random spin rotations \cite{SupplementaryMaterial}. 
At short times, the figure shows that the single-spin subsystem became quickly entangled with the rest of the system, seen as a rapid decrease (increase) of the single-spin purity (entropy), up until the reduced state became completely mixed. At longer times, the purity (entropy) of larger subsystems continued to decrease (increase), as they became entangled with the rest. The dotted curves represent numerical simulations for the experimental parameters, including decoherence, during state initialization, evolution and measurement \cite{SupplementaryMaterial}. 
While panels (a,b) correspond to a specific set of connected partitions $A$, the data gives access to the purities for all partitions $A$ of the system; represented in panel (c) for a specific time $t=5$ ms. Since the second-order R\'enyi entropy of every subsystem is, within three standard deviations, larger than for the total system, this demonstrates entanglement between all $2^{9}-1=511$ bipartitions of the $10$-qubit system.

\begin{figure}
    \centering
    \includegraphics[width=.5\columnwidth]{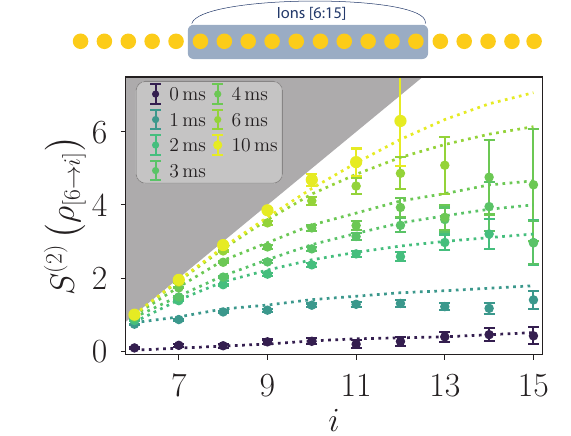}
    \caption{{\bf Second-order R\'enyi entropy of 1- to 10-qubit partitions of a 20-qubit system.} The intial low-entropy N\'eel state evolves under $H_{\mathrm{XY}}$ ($J_0=370~s^{-1}$, $\alpha=1.01$) within 10~ms into a state with high-entropy partitions, corresponding to nearly fully mixed subsystems.
    For the data taken at 6~ms (10~ms) time evolution, the two (three) data points corresponding to highly mixed states are not shown due to their large statistical error bars. For details regarding numerical simulations (dotted curves) and error bars, see~\cite{SupplementaryMaterial}.}
    \label{fig:20-qubit-entropy results}
\end{figure}

Next, a 20-qubit experiment was performed, in which the entropy growth of the central part of the chain was measured during time evolution under $H_{\mathrm{XY}}$, for partitions of up to 10 qubits.  Our observations, shown in Fig.~\ref{fig:20-qubit-entropy results}, are consistent with the formation of highly entangled states. The entropy is seen to increase rapidly over the time evolution of 10~ms, with the reduced density matrices of up to 7 qubits becoming nearly fully mixed. The experimental data agree very well with numerical simulations (dotted curves) obtained with a matrix-product state (MPS) algorithm~\cite{Zaletel:2015}, which includes the (weak) effect of decoherence using quantum trajectories~\cite{Daley:2014}.
The measurement highlights thus the ability of our protocol to access the entropy of highly mixed states, despite larger statistical errors compared to pure states \cite{SupplementaryMaterial}.

Monitoring the entropy growth of arbitrary, yet highly entangled, states during their time evolution constitutes a universal tool for studying dynamical properties of quantum many-body systems, in connection with the concept of quantum thermalization~\cite{Calabrese:2005}. In this context, the entropy growth rate is a decisive quantity to distinguish between the two opposing poles of thermalization and localization in interacting many-body quantum systems \cite{Abanin:2018}.
Generically, in interacting quantum systems without disorder, a ballistic (linear) entropy growth is predicted following a quantum quench~\cite{Calabrese:2005}. Such growth is assumed to persist until saturation is reached, signaling thermalization of the system at late times. On the contrary, in the presence of (strong) disorder and sufficiently short-ranged interactions the  existence of the many-body localized (MBL) phase~\cite{Basko:2006} is predicted in one-dimensional systems \cite{Burin:2015}. This phase is characterized by the absence of thermalization, the system's remembrance on the initial state~\cite{Schreiber:2015} at late times and, in particular, a logarithmic entropy growth~\cite{Bardarson:2012,Serbyn:2013} which constitutes the  distinguishing feature between a MBL state and a non-interacting Anderson insulator. The first experiments probing this entropy growth have been realized with superconducting qubits using tomography~\cite{Xu:2018}, and ultracold atoms based on full-counting statistics of particle numbers~\cite{Lukin2018}. 
For long-range interacting models the situation is less clear, resulting in an ongoing theoretical debate~\cite{Burin:2015, Abanin:2018,Safavi:2018} and first experimental investigations \cite{Smith:2016} into the persistence and stability of localization in such systems.
The measurement of a long-time entropy growth rate is, however, beyond the present capabilities of our trapped-ion quantum simulator, due to its limited coherence time.
As a first application in this context, opening the pathway for future experiments, we now present an observation of the strong diminishing effect of local, random disorder on the entropy growth rate at early times, and the emergence of localization, indicated by a decay of correlations in space and remembrance of the initial state.

\begin{figure}
    \centering
    \includegraphics[width=.7\columnwidth]{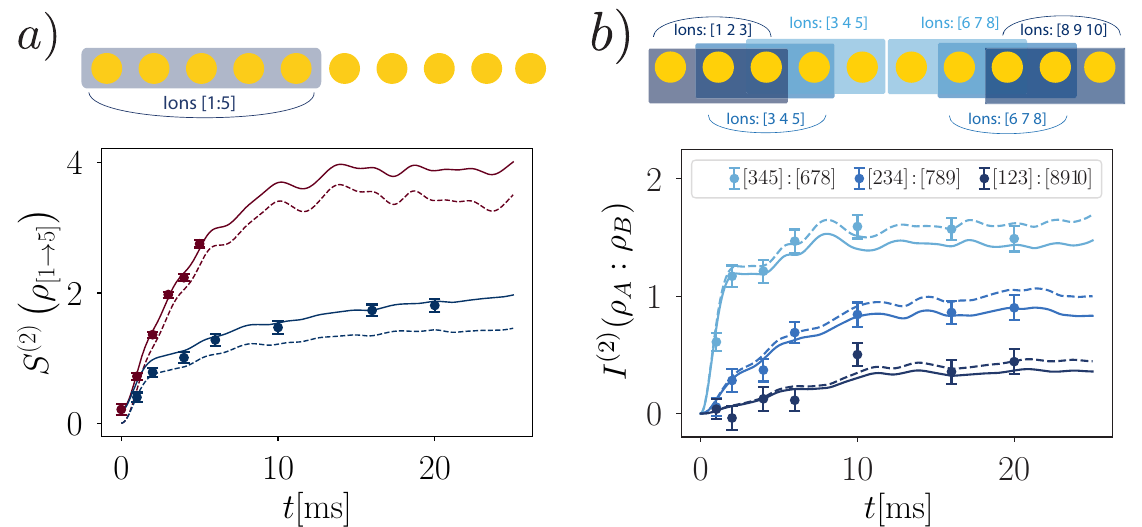}
    \caption{{\bf Spread of quantum correlations under $H_{\mathrm{XY}}$ ($J_0=420\,$s$^{-1}$, $\alpha=1.24$) with and without disorder.} (a) Half-chain entropy growth versus time without disorder (red data points) and with disorder (blue data points). Numerical simulations based on unitary dynamics (dotted curves) including known sources of decoherence (full lines) are in agreement with the measured second-order R\'enyi entropies. (b) Second-order R\'enyi mutual information of selected subsystems versus time. The decrease of $I^{(2)}$ with distance between subsystems is a manifestation of the inhibition of correlation spreading by local disorder. Note that for longer time scales, decoherence leads to a slow increase in the entropy of the total system ($S^{(2)}(\rho) \approx 0.9$ for $t=10$~ms for the full system \cite{SupplementaryMaterial}). Consequently, there is an additional contribution to the slow entropy growth of the system from this decoherence, compared to the case of purely unitary dynamics. Errorbars are the standard error of the mean, caculated with Jacknife resampling of the applied random unitaries.}
    \label{fig:data-with-disorder}
\end{figure}

Fig.~\ref{fig:data-with-disorder}~(a) displays the measured evolution of the second-order R\'enyi entropy at half partition as a function of time, in the absence (presence) of local random disorder. Without disorder, a rapid, linear growth of entropy is observed, in agreement with theoretical simulations including the mentioned sources of decoherence (solid lines).
To investigate the influence of disorder, the initial N\'eel state was quenched with $H=H_{\mathrm{XY}}+H_{\mathrm{D}}$, where the static, random disorder strength $\Delta_j$ was drawn uniformly from $[-3J_0, 3J_0]$. To efficiently access disorder-averaged quantities, our protocol offers the possibility to combine ensemble average over random unitaries and the disorder average \cite{SupplementaryMaterial}. Hence, only $10$ random unitaries per disorder pattern ($N_M=150$ measurements per unitary), and $35$ randomly drawn disorder patterns were used to obtain an accurate estimate of the disorder-averaged purity $\widetilde{\tr{\rho_A^2}}$ ($\widetilde{\dots}$ denotes the disorder average), and subsequently the second-order R\'enyi entropy $\widetilde{S^{(2)}(\rho_A)} \approx -\log_2\widetilde{\tr{\rho_A^2}}$~\cite{SupplementaryMaterial}. The measured, disorder-averaged entropy growth clearly demonstrates how disorder reduces the growth of entanglement. After an initial rapid evolution, a considerable slowing of the dynamics is observed, with a small, but non-vanishing, growth rate at later times; a behaviour compatible with the scenario of MBL. 
This observation is accompanied with a remembrance of the initial N\'eel state during the dynamics, manifest in the measured time evolution of the local magnetization \cite{SupplementaryMaterial}.

Finally, Fig.~\ref{fig:data-with-disorder} (b) shows the evolution of the second-order R\'enyi mutual information, defined as $I^{(2)}(\rho_{A}:\rho_{B}) = S^{(2)}(\rho_{A}) + S^{(2)}(\rho_{B}) - S^{(2)}(\rho_{AB})$, quantifying the total amount of classical and quantum correlations between various pairs of subsystems \cite{Islam:2015}. In the presence of disorder, $I^{(2)}(\rho_{A}:\rho_{B})$ saturates quickly to approximately constant values, which decrease with increasing distance between the two partitions $A$ and $B$. This spatial decay of correlations provides a further indication of localization due to the presence of disorder in our system. 
 
We have demonstrated a new tool for measuring second-order R\'enyi entropies, and shown how it provides a powerful method for both  characterizing engineered quantum systems and using them to tackle open questions in physics. 
In our experiments, we studied the entropy of partitions of up to 10 qubits, due to technical restrictions that currently limit our experimental repetition rate. Straightforward technical improvements should allow the entropy of 20 qubit systems to be measured in our system in the near future. Numerical simulations \cite{SupplementaryMaterial} indicate that the total number of measurements required to access the purity within a statistical error of $0.12$ is, for a pure product state of $N_A$ qubits, given by $2^{7.7\pm 0.3+(0.8\pm 0.1) N_A}$. The amount of measurements required to obtain the purity of entangled pure states can be significantly lower~\cite{SupplementaryMaterial}.
Purity measurements of systems containing tens of qubits is therefore likely also in reach in experiments with high quantum state generation rates, such as state-of-the-art superconducting qubit setups. 
The number of measurements could be further decreased by replacing the local random operations by global random unitaries acting on the entire Hilbert space of a subsystem of interest, via random quenches~\cite{Elben:2018,Vermersch:2018},
at the expense of obtaining access to the purity of a single partition only.




\bibliographystyle{Science}

\section*{Acknowledgments}
We acknowledge funding from the ERC Synergy Grant UQUAM, from the European Research Council (ERC) under the European Union's Horizon 2020 research and innovation programme under grant agreement No 741541, and the SFB FoQuS (FWF Project No. F4016-N23).

\section*{Author contributions}
PZ suggested the research topic, which was further developed by AE, BV, BPL, and CFR.  AE, BV and PZ developed the theoretical protocols. PJ, CM, TB, BPL, CFR, and RB contributed to the experimental setup. TB, PJ, CM, and CFR performed the experiments. AE, BV and CFR analyzed the data and carried out numerical simulations. TB, AE, BV, BPL, PZ, and CFR wrote the manuscript. All authors contributed to the discussion of the results and the manuscript.

\section*{Supplementary materials and methods}
\setcounter{section}{1}

\subsection{Purity from randomized measurements}

In this section, we derive  Eq.~(2) of the main text which relates the second order moments of outcome probabilities of randomized measurements on a subsystem $A$, with the purity $\tr[]{\rho_A^2}$ of its reduced density matrix $\rho_A$.
Generalizing the treatment in the main text, we assume that $A$ is composed of $N_A$ constituents with arbitrary Hilbert space dimension $d$.   

The purity $\tr[]{\rho_A^2}$ of the reduced density matrix $\rho_A=\tr[\mathcal{S}/A]{\rho}$ of a subsystem $A$ in the total system $\mathcal{S}$, consisting of $N_A$  constituents $(i[1],\dots,i[{N_A}])$, is  inferred from randomized measurements performed on $A$. 
These are  implemented using random unitaries given as
\begin{align}
U_A= \bigotimes_{l=1}^{N_A} u^{(i[l])}  \; ,
\end{align}
where  each $u^{(i[l])} \in \text{CUE}(d) $,  acting on the local constituent $i[l]$, is drawn independently from the circular unitary ensemble (CUE). Subsequently, a measurement in a fixed basis is performed.  We assume this basis to be given by the product states $\left\{  \ket{\mathbf{s}_A}  \right\} = \left\{  \ket{s_{i[1]}, \dots, s_{i[{N_A}]}}  \right\} $ where $s_{i[l]}\in \{ 1,\dots,d \}$ labels the basis states in the Hilbert space of constituent $i[l]$.  The probability to obtain from such a measurement an outcome $\mathbf{s}_A=(s_{i[1]},\dots,s_{i[N_A]})$ is then given by $P(\mathbf{s}_{A}) =\tr[A]{U_A \rho_A U_A^\dagger \ket{\mathbf{s}_A}\bra{\mathbf{s}_A}}$. 

In the following, we show  that the purity $\tr[]{\rho_A^2}$ is  obtained from the ensemble average  of cross correlations $\overline{\probsU{\mathbf{s}_{A}}\probsU{ \widetilde{\mathbf{s}_{A}}}  }  $  over the random unitaries via
\begin{align}
\tr[]{\rho_A^2} = d^{N_A}\sum_{\mathbf{s}_{A}\widetilde{\mathbf{s}_{A}} } (-d)^{-D[\mathbf{s}_{A},\widetilde{\mathbf{s}_{A}}]} \; \overline{\probsU{\mathbf{s}_{A}}\probsU{ \widetilde{\mathbf{s}_{A}}}  }    \label{eq: trrho2} \;.
\end{align}
Here, the sum extends over all basis states $\mathbf{s}_{A}$, $ \widetilde{\mathbf{s}_{A}}$. The Hamming distance $D[\mathbf{s}_{A},\widetilde{\mathbf{s}_{A}}]$  between  two  states $\mathbf{s}_{A}$ and $ \widetilde{\mathbf{s}_{A}}$ is defined as  the number of local constituents $i\in A$    where $s_{i} \neq \widetilde{s}_{i}$, i.e.\  $D[\mathbf{s}_{A},\widetilde{\mathbf{s}_{A}}]\equiv  \# \left\{ i \in A  \, | \, s_{i} \neq \widetilde{s}_{i} \right\}$.

We note that, in practise, we use unbiased estimators~($\mathit{22}$) of the cross correlations $\probsU{\mathbf{s}_{A}}\probsU{ \widetilde{\mathbf{s}_{A}}}$ for a subsystem $A$, to estimate them faithfully from a finite number of projective measurements.
We further remark that Eq.~(\ref{eq: trrho2}) represents, in contrast to the recursive scheme presented in Ref.~($\mathit{21}$), an explicit formula to reconstruct the purity of an arbitrary subsystem $A$. It  can be used to access the purity of all subsystems of interest simultaneously, after performing randomized measurements on the entire system $\mathcal{S}$ (see Fig.\ 2 of the main text).
Moreover, we note that, restricted to the case of a single constituent $N_A=1$ in $A$ of dimension $d$, Eq.~(\ref{eq: trrho2}) represents an alternative way to extract the purity from statistical correlations of randomized measurements implemented with global random unitaries ($\mathit{20}$,$\mathit{21}$).

In the following, we prove Eq.~(\ref{eq: trrho2}). To ease notation, we assume  without loss of generality that the constituents in $A$ are labeled consecutively, i.e.\ $(i[1],\dots,i[{N_A}])=(1,\dots, N_A)$. Key ingredient in the proof are the $2$-design properties ($\mathit{20}$,$\mathit{22}$) of the local random unitaries $u^{(i)} \in \text{CUE}$:  
\begin{align}
\overline{ u^{(i)}_{s_i,s'_i} \left(u^{(i)}_{s_i,s''_i}\right)^*u^{(i)}_{\tilde{s}_i,\tilde{s}'_i} \left(u^{(i)}_{\tilde{s}_i,\tilde{s}''_i}\right)^*   } = &  \frac{\delta_{s'_i,s_i''}\delta_{\tilde{s}'_i,\tilde{s}_i''}+\delta_{s_i,\tilde{s}_i}\delta_{s'_i,\tilde{s}_i''}\delta_{\tilde{s}_i',s''_i}}{d^2-1}-\frac{\delta_{s'_i,\tilde{s}_i''}\delta_{\tilde{s}'_i,s_i''}+\delta_{s_i,\tilde{s}_i}\delta_{s'_i,s_i''}\delta_{\tilde{s}'_i,\tilde{s}_i''}}{d(d^2-1)} \label{eq:2design} 
\end{align}
Here, as before, $\overline{\vphantom{P}\dots}$ denotes the ensemble average.
We further note that  the probabilities $P(\mathbf{s}_{A}) $ are given by
\begin{align}
    P(\mathbf{s}_{A})=\tr[]{U_A \rho_A U_A^\dagger \ket{\mathbf{s}_A}\bra{\mathbf{s}_A}} = \prod_{i \in A}  u^{(i)}_{s_{i}, s_{i}'} \left( u^{(i) }_{s_{i}, s_{i}''} \right)^*  (\rho_A) _{(s_{1}', \dots, s_{N_A}') (s_{1}'', \dots, s_{N_A}'')} \, ,
\end{align}
where here, and in the following, summation over primed indices is implied. 
Using the independence of the local random unitaries $u^{(i)}$, we then find
	\begin{align}
		\overline{\probsU{\mathbf{s}_{A}}\probsU{ \widetilde{\mathbf{s}_{A}}}  } = &\prod_{i \in A} \overline{ u^{(i)}_{s_i,s'_i} \left(u^{(i)}_{s_i,s''_i}\right)^*u^{(i)}_{\tilde{s}_i,\tilde{s}'_i} \left(u^{(i)}_{\tilde{s}_i,\tilde{s}''_i}\right)^*   } \nonumber \\
		& \qquad \times (\rho_A) _{(s_1', \dots, s_{N_A}') (s_1'', \dots, s_{N_A}'')}    (\rho_A) _{(\tilde s_1', \dots, \tilde s_{N_A}') (\tilde s_1'', \dots,\tilde s_{N_A}'')}   
\nonumber \\
= & \prod_{\substack{i \in A \\ s_i = \tilde{s}_i}} \frac{\delta_{s_i' s_i''}\delta_{\tilde s_i' \tilde s_i''}+\delta_{s_i'  \tilde s_i''}\delta_{\tilde s_i'  s_i''}}{d(d+1)} \prod_{\substack{i \in A \\ s_i \neq \tilde{s}_i}} \frac{\delta_{s_i' s_i''}\delta_{\tilde s_i' \tilde s_i''}-\frac{1}{d}\delta_{s_i'  \tilde s_i''}\delta_{\tilde s_i'  s_i''}}{d^2-1} \nonumber \\ 
& \qquad \times (\rho_A) _{(s_1', \dots, s_{N_A}') (s_1'', \dots, s_{N_A}'')}    (\rho_A) _{(\tilde s_1', \dots, \tilde s_{N_A}') (\tilde s_1'', \dots,\tilde s_{N_A}'')}   
\nonumber \\
=& \frac{\sum_{A'\subseteq A} (-d)^{-D[\mathbf{s}_{A'},\widetilde{\mathbf{s}_{A'}}]} \tr[]{\rho_{A'}^2}}{(d(d+1))^{N_A-D[\mathbf{s}_{A},\widetilde{\mathbf{s}_{A}}]}(d^2-1)^{D[\mathbf{s}_{A},\widetilde{\mathbf{s}_{A}}]}} \label{eq:pkk}
	\end{align}
where in the  last line, we sum over all subsystems $A'\subseteq A$, including the empty subset for which we define $\tr[]{\rho_\emptyset^2}\equiv 1$. Further, $\mathbf{s}_{A'}\equiv\mathbf{s}_{A}|_{A'}$ and  $\widetilde{\mathbf{s}_{A'}}\equiv\widetilde{\mathbf{s}_{A}}|_{A'}$ denote the restrictions of $\mathbf{s}_A$ and $\widetilde{\mathbf{s}_A}$  to $A'$ such that $D[\mathbf{s}_{A'},\widetilde{\mathbf{s}_{A'}}]=   \# \left\{ i \in A'\subseteq A  \, | \, s_{i} \neq \widetilde{s}_{i} \right\}$.
Inserting Eq.~\eqref{eq:pkk} into Eq.~\eqref{eq: trrho2} we  obtain 
\begin{align}
	d^{N_A}\sum_{\mathbf{s}_{A}\widetilde{\mathbf{s}_{A}}} (-d)^{-D[\mathbf{s}_{A},\widetilde{\mathbf{s}_{A}}]}   \;  \overline{\probsU{\mathbf{s}_{A}}\probsU{ \widetilde{\mathbf{s}_{A}}}  }  = \sum_{A' \subseteq A}   C_{A'} \tr[]{\rho_{A'}^2} \; 
\end{align}
with
\begin{align*}
C_{A'}= d^{N_A} \sum_{\mathbf{s}_{A}\widetilde{\mathbf{s}_{A}}} \frac{ (-d)^{-D[\mathbf{s}_{A},\widetilde{\mathbf{s}_{A}}]-D[\mathbf{s}_{A'},\widetilde{\mathbf{s}_{A'}}]}}{ (d(d+1))^{N_A-D[\mathbf{s}_{A},\widetilde{\mathbf{s}_{A}}]}(d^2-1)^{D[\mathbf{s}_{A},\widetilde{\mathbf{s}_{A}}]}} \; .
\end{align*}
Selecting an arbitrary basis state $\mathbf{s}_A$, we find after some manipulation
\begin{align*}
C_{A'}&= \left( \frac{d}{d+1}\right)^{N_A} \sum_{\widetilde{\mathbf{s}_{A}}} \; \frac{(-1)^{D[\mathbf{s}_{A},\widetilde{\mathbf{s}_{A}}] - D[\mathbf{s}_{A'},\widetilde{\mathbf{s}_{A'}}]}}{ d^{D[\mathbf{s}_{A'},\widetilde{\mathbf{s}_{A'}}]} (d-1)^{D[\mathbf{s}_{A},\widetilde{\mathbf{s}_{A}}]}} \\
&= \left( \frac{d}{d+1}\right)^{N_A}  \sum_{m=0}^{N_{A'}} \binom{N_{A'}}{m} (d-1)^m \; \sum_{l=0}^{N_A-N_{A'}} \binom{N_A-N_{A'}}{l} (d-1)^l  \nonumber \\  & \qquad \qquad \qquad \qquad \times \frac{(-1)^l}{d^m (d-1)^{m+l}}    \nonumber \\
&= \left( \frac{d}{d+1}\right)^{N_A}  \underbrace{ \sum_{m=0}^{N_{A'}} \binom{N_{A'}}{m}  \frac{1}{d^m}}_{=  \left( \frac{d+1}{d}\right)^{N_{A'}} } \underbrace{\sum_{l=0}^{N_A-N_{A'}} \binom{N_A-N_{A'}}{l} (-1)^l}_{\delta_{N_A N_{A'}}} \nonumber \\
&= \delta_{N_A N_{A'}} \;. 
\end{align*}
From first to second line, we used that, for $m\in \{0,\dots,N_{A'}\}$,  there are $\binom{N_{A'}}{m} (d-1)^m$ states in the basis of $A'$ with $D[\mathbf{s}_{A'},\widetilde{\mathbf{s}_{A'}}]=m$. Equally, for the complement $A \backslash  A' $, there are, for $l\in \{ 0,\dots,N_A-N_{A'}\}$, $\binom{N_A-N_{A'}}{l} (d-1)^l$ states in the basis of $A \backslash  A' $ with $D[\mathbf{s}_{A \backslash  A'},\widetilde{\mathbf{s}_{A \backslash  A'}}]=l$.
As $A'\subseteq A$, so follows the claim.

\subsection{Scaling of the required number of measurements}

In this section, we first discuss the scaling of the required number of measurements to determine the purity (second-order R\'enyi entropy) of a single density matrix $\rho_A$ up to a fixed relative (absolute) statistical error with (sub-)system size $N_A$. 
Secondly, we show that in the context of disordered systems, the disorder-averaged purity (disorder-averaged second-order R\'enyi entropy) can be accessed efficiently, i.e.\ without further increase of the required number of measurements compared to a single purity estimation in a clean system.

\subsubsection{Single quantum states}

The total number of measurements required to estimate the purity of a given reduced density matrix, $\rho_A$, is  given  as the product of the number $N_U$ of random unitaries applied, with $U=u_1 \otimes \dots \otimes u_{N_A}$,  and the number $N_M$ of projective measurements per random unitary. In order to minimize the total number of measurements, $N_U N_M$, while keeping the statistical error below a predefined threshold, an optimal ratio $N_U/N_M$ must be chosen, which, as we discuss below, depends on the quantum state of interest. We note already here, however, that the pure product states are most prone to statistical errors, i.e. the \emph{absolute} statistical error of a purity estimation with a given set of $N_U$ unitaries and $N_M$ measurements is, for an arbitrary (entangled or mixed) quantum state, lower than for a pure product state.  This is explained by the fact that, for  a mixed or entangled state whose subsystems are  mixed, fluctuations across  the unitary ensemble are reduced (compared to the pure product state). This leads  to a smaller statistical error (see also Refs.~\cite{vanEnk:2012,Vermersch:2018}).

To access the statistical error of a purity estimation  to be expected on average in an experiment, we performed, for a given quantum state $\rho_A$ of $N_A$ qubits and a given number of unitaries $N_U$ and measurements $N_M$, $100$ numerical simulations of an experiment. Here, the individual single qubit random unitaries, $u_i$, were independently sampled from the CUE using the algorithm given in Ref.~\cite{Mezzadri:2007}.  From the set of estimated purities, $\left( \tr[]{\rho^2_A} \right)_e$, we determined the average absolute  $ | \left( \tr[]{\rho^2_A} \right)_e  -  \tr[]{\rho^2_A}| $ and relative $ \Delta_e= \Delta_e(\rho_A,N_U,N_M)\equiv  | \left( \tr[]{\rho^2_A} \right)_e  -  \tr[]{\rho^2_A}|/\tr[]{\rho^2_A}$ error. From  the relative error, $\Delta_e$,  one determines directly the absolute error of an estimation $ \left( S^{(2)}(\rho_A)  \right)_e = - \log_2 \left( \tr[]{\rho^2_A} \right)_e$ of the second-order R\'enyi entropy, $  |\left(  S^{(2)}(\rho_A)  \right)_e -  S^{(2)}(\rho_A)| = \Delta_e \ln 2  + \mathcal{O} \left( \Delta_e^2 \right)$.
Finally, to find the optimal ratio $N_U/N_M$, which minimizes the total number of measurements, $N_U N_M$, while keeping the average relative error, $\Delta_e$,  below $12 \%$, we varied $N_U$ and $N_M$ on a $50 \times 50$ optimization grid with range $N_U=4,\dots, 1024$ (quadratic spacing) and $N_M=4,\dots, 1024$ (logarithmic spacing). 

\begin{figure}
    \centering
    \includegraphics[width=1\columnwidth]{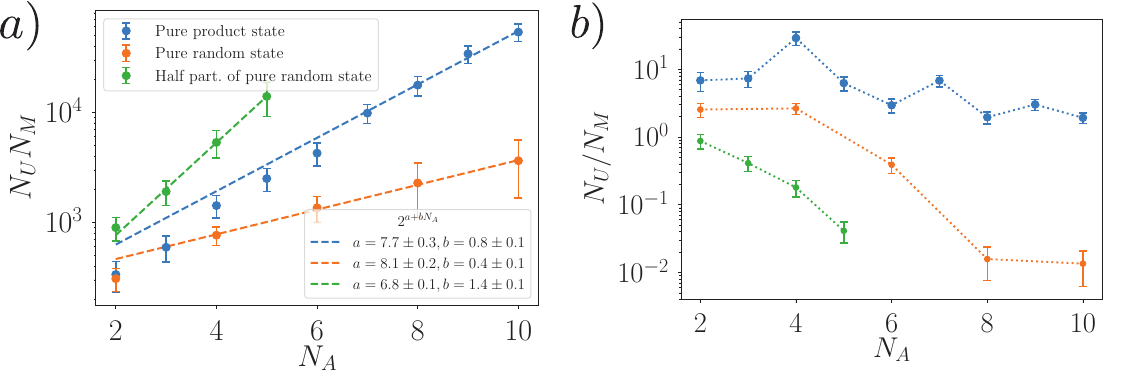}
    \caption{{\bf Scaling of the required number of measurements.} The required number of measurements $N_U N_M$ [panel $(a)$] to determine the purity up to an average relative error of $0.12$ scales exponentially with subsystem size $N_A$. The exponents of the exponential fits, depending on the state of interest (see text) are favorable compared to tomography \cite{Gross:2010}. Here, the ratio $N_U/N_M$ [panel $(b)$] has been optimized numerically; error bars in both panels are due to the finite resolution of the optimization grid. }
    \label{fig:scaling}
\end{figure}

In Fig.~\ref{fig:scaling},  the results of this optimization  are displayed. In panel $(a)$, the required number of measurements $N_U N_M$ is shown, for the limiting cases of  pure product states (blue), pure highly entangled random states (orange), obtained by applying a random unitary from the CUE to a product state, and highly mixed states (green), obtained from the reduced density matrix at half partition of  Haar-random states. We find that the total number of measurements scales exponentially with subsystem size $N_A$. However, the corresponding exponents are, compared to full state tomography, favorable \cite{Gross:2010}. Since we consider the relative error of the estimated purity, the highly mixed state has the largest exponent of $1.4 \pm 0.1$. For the pure product state, the exponent is $0.8 \pm 0.1$, whereas for a pure Haar-random state it is further decreased to $0.4 \pm 0.1$. 
To employ this reduction of the number of measurements,  the ratio $N_U/N_M$ [Fig.~\ref{fig:scaling} panel (b)] has to be chosen accordingly. For an entangled and/or mixed state, a ratio $N_U/N_M\ll1 $ is optimal, in contrast to the pure product state $N_U/N_M>1$.
This  is explained by the fact that, for a mixed and/or entangled state whose subsystems are highly mixed, fluctuations across the unitary ensemble are reduced. Thus, a smaller number of unitaries is sufficient to determine the purity up to a given absolute error. For the pure random state, this results in a smaller total number of measurements for a determination of the purity up to a relative error below $12 \%$. For the highly mixed state, the purity itself is small, and thus the relative error, i.e. the total number of measurements, is increased.

We note that, with our present scheme, the choice of the optimal ratio $N_U/N_M$ would require a priori knowledge of the state of interest; we expect, however, that adaptive (re-)sampling schemes have the potential to determine the optimal ratio $N_U/N_M$ based on experimental data in the future. Additionally, as mentioned previously, for a given, predefined choice $N_U$ and $N_M$, the \emph{absolute} statistical error of the estimated purity of an entangled state and/or mixed is always lower than for a  product state. In particular, and in contrast to recent tomographic approaches based on a particular variational ansatz \cite{Lanyon:2017,Torlai:2018}, our protocol performs hence well on random or very chaotic states with highly mixed subsystems. Despite growing \emph{relative} statistical errors, this allows one to measure purities (second order R\'enyi entropies) of highly mixed states; as demonstrated in Fig. 3 (main text).

We remark further that  statistical errors of the estimated purity (second order R\'enyi entropy), being the standard error of the mean of $X$ (Eq.\ 2 main text), can be determined from the experimental data itself. The accuracy of an estimation is thus always determined from the experimental data, without any assumption on the quantum state of interest.

\subsubsection{Disorder-averaged purity}

 In this section, we numerically show that our protocol can be used to  access the disorder-averaged purity (second-order R\'enyi entropy) in an efficient way: by combining disorder and ensemble averaged over random unitaries, one can estimate disorder averaged quantities without increasing the total number of measurements, compared to the purity estimation in a clean system. The protocol is hence optimally suited for the study of entropy dynamics  in disordered systems.

\begin{figure}
    \centering
    \includegraphics[width=1\columnwidth]{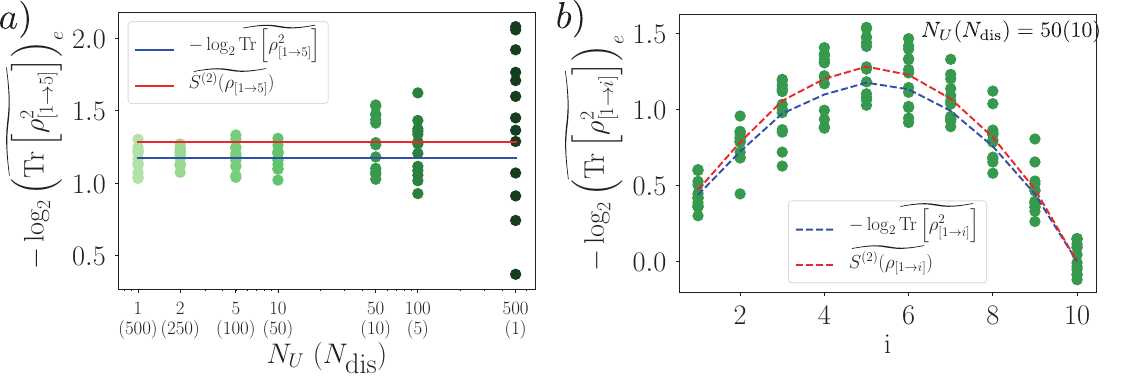}
    \caption{{\bf Efficient merging of disorder and random unitary average.} In panel $(a)$, the negative logarithm of estimations of the disorder-averaged half-chain purity (see text) are displayed for various combinations of the number of disorder realizations $N_{\text{dis}}$ and number of unitaries $N_U=N_{\text{dis}}/500$ (green dots). In panel $(b)$, for a given pair $N_{\text{dis}}=10, N_U=50$, the negative logarithm of the estimated disorder averaged purities is shown, for various subsystems $[1\rightarrow i]$, $i=1,\dots,10$. The red (blue) line corresponds in both panels to the negative logarithm of the exact  purity (exact second-order R\'enyi entropy) averaged over $500$ disorder patterns.  }
    \label{fig:disav}
\end{figure}

To demonstrate this numerically, we randomly sample $500$ disorder patterns $\Delta_j^i \in \left[ - 3J_0, 3 J_0\right]$ (disorder index $i=1,\dots,500$ and spatial index $j=1,\dots, 10$) and evolve via exact diagonalization an initial 10-qubit Neel state under the Hamiltonians $H^i=H_{\text{XY}}+\hbar \sum_j \Delta_j^i \sigma_j^z$ to a time $t_f=25$~ms, to obtain an ensemble of states $\left\{\rho^i(t_f) \right\}_{i=1,\dots,500}$.
Subsequently, the disorder-averaged purity $\widetilde{\tr[]{\rho^2}}\equiv 1/500 \sum _i {\tr[]{(\rho_A^{(i)})^2}}$ and the disorder-averaged second-order R\'enyi entropy $\widetilde{S^{(2)}(\rho_A)}\equiv 1/500 \sum _i {S^{(2)}(\rho_A^{(i)})}$ are calculated, where here $\rho^{(i)}_A=\tr[A^c]{\rho^{(i)}(t_f)}$. Both quantities are displayed in Fig.~\ref{fig:disav} as blue and red lines respectively. Due to the concavity of the logarithm, $-\log_2 \widetilde{\tr[]{\rho_A^2}}$ underestimates the disorder-averaged second-order R\'enyi entropy $\widetilde{S^{(2)}(\rho_A)}$. However, this bias is, for the small entropies typically present in disordered systems, in the range of a few percent.

To show that disorder and random unitary average can be combined efficiently, we randomly select $N_{\text{dis}}$ states from $\left\{\rho^i(t_f) \right\}_{i=1,\dots,500}$ and use $N_U=500/N_{\text{dis}}$ random unitaries to perform a numerical simulation of an experiment according to our protocol (no projection noise included). From this, we obtain an estimation of the disorder averaged purity $\left(\widetilde{\tr[]{\rho_A^2}} \right)_e$.
The results of $13$ numerical experiments are displayed in Fig.~\ref{fig:disav} for various combinations $(N_{\text{dis}}, N_U=500/N_{\text{dis}})$ (green dots). As is clearly visible, by choosing many different disorder patterns $N_{\text{dis}}=500, \dots, 10$, but only very few random unitaries ($N_U=1,\dots,50$), one obtains a precise and faithful estimation $\left(\widetilde{\tr[]{\rho_A^2}} \right)_e$ of the disorder-averaged purity $\widetilde{\tr[]{\rho_A^2}}$. From this, we obtain with $- \log_2 \left(\widetilde{\tr[]{\rho_A^2}} \right)_e$ an estimation of $\widetilde{S^{(2)}(\rho_A)}$ whose bias is,  for the values of $N_{\text{dis}}$, $N_U$ and $N_M$ used in the experiment, within (statistical) error bars.

\subsection{Experimental setup and implementation}

The trapped-ion platform was realized using a string of $^{40}\mathrm{Ca}^{+}$ ions confined in a linear Paul trap. Each qubit was encoded in the Zeeman sublevels of two long-lived internal electronic states of each ion, the $S$ and $D$ states. The computational basis states of the qubit were chosen to be $|\!\downarrow\rangle = |S_{1/2},m_{j}=1/2\rangle$ and $|\!\uparrow\rangle = |D_{5/2},m_{j}=5/2\rangle$. These states are coupled through an optical quadrupole transition with transition frequency $\omega_0/(2\pi)\approx 411$~THz. The transition is driven using a global beam (i.e. illuminating all ions simultaneously) coupling the two qubit states, and an off-resonant, tightly-focused, single-ion addressed laser beam that can be steered with an acousto-optical deflector. Both beams are perpendicular to the ion string in order to avoid coupling to the Doppler-cooled axial modes of motion of the ion string. The degeneracy of the Zeeman states is lifted by using a magnetic field of 4.18~G, allowing optical pumping to the $|S_{1/2},m_{j}=1/2\rangle$ state with approximately 99.9\% efficiency \cite{Jurcevic:2014}. Doppler cooling followed by sideband cooling on the quadrupole transition prepares all transverse collective modes close to their ground states. A combination of global and single-ion addressed beams initialized the N\'eel-ordered product state $|\!\uparrow\downarrow\uparrow...\rangle$ using composite global pulses (see Supplementary Information of Ref.~\cite{Lanyon:2017}). Decoherence by laser phase and intensity noise led to imperfect preparation of the N\'eel-ordered state, resulting in a purity loss of approximately 0.08 for a 10-qubit N\'eel state and 0.19 for a 20-qubit N\'eel state.

Spin-spin Ising-type interactions were realized through a laser beam that off-resonantly coupled all ions on the qubit transition to all transverse collective motional modes  of the ion string,
realizing the effective Hamiltonian
\begin{equation}\label{Ising}
H_{\mathrm{Ising}} = \hbar \sum_{i<j}J_{ij}\sigma_{i}^{x}\sigma_{j}^{x}+\hbar B\sum_{i}\sigma_{i}^{z}
\end{equation}
where $\sigma_{i}^{\beta}$, with $\beta=\{x,y,z\}$, are the Pauli spin-1/2 matrices for the $i$th spin, and $B$ is the effective transverse magnetic field strength. The qubit-qubit interactions, $J_{ij} \approx J_0/|i-j|^{\alpha}$, follow an approximately power-law dependence with distance $|i-j|$. 
For the experiments conducted with strings of 10 ions, $\alpha \approx1.24$ and $J_0 \approx 420\,\mathrm{s}^{-1}$ (the value of $\alpha$ was inferred from the dispersion relation as described in Ref.~\cite{Jurcevic:2014}). For those experiments with 20 ions, $\alpha\approx 1.01$ and $J_{0} \approx 370\,\mathrm{s}^{-1}$. 
The Hamiltonian of Eq.~(\ref{Ising}) was implemented using a bichromatic laser beam carrying frequencies $\omega_{\pm}=\omega_{0}\pm\Delta$, where $\Delta$ was chosen such that the bichromatic beams were detuned by $\pm 40$~kHz from the first-order sideband transitions of the collective mode of motion with the highest frequency. The transverse field component, $\hbar B\sum_{i}\sigma_{i}^{z}$, was implemented by introducing an additional detuning of $\delta/(2\pi)=3$~kHz, such that the bichromate frequencies were $\omega_{\pm}=(\omega_{0}\pm\Delta) + \delta$. This value of $\delta$ was in the regime where $B \gg J_{0}$, having the desirable characteristic of conserving the number of spin excitations, $\ket{\uparrow}$, throughout the dynamics. Consequently $H_{\mathrm{Ising}}$ reduced to the XY interaction Hamiltonian as given in Eq.~(\ref{eq:XY-Hamiltonian}) of the main text.\\

\noindent\textbf{Generation of random disorder potentials}\\
Random disorder potentials were implemented through local $\sigma_{i}^{z}$ rotations on all ions. A detuned laser beam  at 729~nm, deflected from an acousto-optical deflector (AOD), generated independently controllable AC-Stark shifts on all ions simultaneously \cite{Maier:2018}. These AC-Stark shifts implemented a time-independent alteration to the transverse field. Tuning the power of the 729nm beam incident on each ion allowed independent tuning of the disorder on each ion, in the range of $\Delta_j=[0, 6J_0]$.

\subsection{Randomized measurements}
\textbf{Generation of random unitaries}\\
The measurement protocol consisted of applying to each qubit a local random unitary matrix
\begin{equation}
U=\begin{bmatrix}
U_{1} & U_{2}\\
-U_{2}^{*} & U_{1}^{*}
\label{eq:RandomUnitary}
\end{bmatrix}
\end{equation}
drawn from the circular unitary ensemble (CUE) \cite{Elben:2018,Vermersch:2018}, following the algorithm given in Ref. \cite{Mezzadri:2007}. Note that in Eq.~(\ref{eq:RandomUnitary}) we have dropped a global phase factor.
The distribution of random unitaries act such that any state on the surface of the Bloch sphere, i.e. any pure state, has an equal probability of being rotated into any other state on the surface of the Bloch sphere.
Any such unitary operator can be written as a combination of rotations, and so can be decomposed into rotation angles around the X, Y and Z axes of the Bloch sphere. One possible decomposition consists in searching for those real numbers $\theta_{1}$, $\theta_{2}$ and $\theta_{3}$ such that
\begin{equation}
U = R_{z}(\theta_{3})R_{y}(\theta_{2})R_{z}(\theta_{1})
\label{eq:threepulsedecomposition}
\end{equation}
where $R_{\beta}(\theta) = e^{-i\sigma^{\beta}\theta/2}$, $\sigma^{\beta}$ are the Pauli matrices with $\beta =\{ x,y,z\}$ and $\theta$ are the rotational angle. A straightforward calculation shows that the rotation angles are related to the matrix elements of $U$ via \begin{align*}
    \theta_{1} &= \phantom{2}\Re{\left\{\mathrm{tan}^{-1} \left[ \frac{-i(U_{1}^{*}-U_{1})}{U_{1}^{*}+U_{1}} \right] - \mathrm{tan}^{-1} \left[ \frac{-i(U_{2}^{*}-U_{2})}{U_{2}^{*}+U_{2}} \right]\right\} }\\
    \theta_{2} &= \phantom{2}\Re{\left\{2\,\mathrm{tan}^{-1}\left[\left(\frac{|U_{2}|^2}{|U_{1}|^2}\right)^{1/2}\right]\right\}}\\
    \theta_{3} &= \phantom{2}\Re{\left\{\mathrm{tan}^{-1} \left[ \frac{-i(U_{1}^{*}-U_{1})}{U_{1}^{*}+U_{1}} \right] + \mathrm{tan}^{-1} \left[ \frac{-i(U_{2}^{*}-U_{2})}{U_{2}^{*}+U_{2}} \right]\right\}}.
\end{align*}

\vspace{.5cm}
\noindent \textbf{Characterization of local random unitaries}\\ 
For a single qubit prepared in a pure state, application of a random unitary drawn from the CUE should take the Bloch vector describing the qubit state to any point on the Bloch sphere with equal likelihood. As a consequence, the distribution of measured spin projections should be uniform along any direction of projection.

However, systematic errors, such as minor miscalibrations of global $\pi/2$-pulses, will affect the distribution of random unitaries, which can give rise to erroneous (or even unphysical) purity values when using the random measurement protocol for the determination of the purity of a given qubit state. A second source of errors that can bias the purity estimates towards lower values is decoherence occuring during the application of the local random unitaries. Both of these effects can be detected by an analysis of the distribution of measured spin projections for a qubit prepared in a pure state.

To make the drawing of random unitaries from the CUE more robust against miscalibration or drift of experimental control parameters, we concatenated two random unitaries (drawn from a possibly imperfect distribution) to obtain a random unitary with a distribution that is closer to the ideal one. Instead of concatenating twice the pulses given in (\ref{eq:threepulsedecomposition}), we implemented the random unitary  in the experiment by sandwiching addressed light-shift pulses between global $\pi/2$ pulses realizing the local unitaries
\begin{equation}
U = R_x(-\pi/2)R_{z}(\theta_{4})R_x(\pi/2)\,
    R_{z}(\theta_{3})\,
    R_x(-\pi/2)R_{z}(\theta_{2})R_x(\pi/2)\,
    R_{z}(\theta_{1})
    \label{eq:experimental-pulse-sequence}
\end{equation}
where the $\pi/2$ pulses serve to convert the required addressed $y$-rotations into $z$-rotations, since $z$-rotations are the only rotations that can be carried out with our addressed beam. Additionally, we merged the two consecutive rotations around $z$ into a single rotation and dropped the final $z$-rotation pulse that would not have affected the measurement results (as the measurements were done in the $z$-basis).
In this way, by replacing the rotations $R_z(\theta)$ in (\ref{eq:experimental-pulse-sequence}) by blocks of addressed rotations, $\prod_{i=1}^N R_z^i(\theta_i)$, the desired tensor product of local random unitaries could be realized. As we could realize only $z$-rotations with a positive rotation angle by ac-Stark shifting the energy levels of the ions, we replaced all negative rotations angles $\theta$ by $2\pi-\theta$. Moreover, we minimized the duration of the pulse length by replacing $\theta\rightarrow\tilde\theta=\mbox{mod}(\theta_i-\alpha,2\pi)$, where $\alpha$ minimizes the function $\sum_i\tilde{\theta}$, and shifting the rotation axis of subsequent resonant global pulses in the equatorial plane accordingly by an angle $\alpha$.

Concatenation of random unitaries leads to an improved robustness of the realized distribution, but comes at the price of increased decoherence.
To investigate these effects, we prepared a 10-ion string 
by optical pumping in a nearly perfectly pure state before applying the laser pulses realizing the (double) random unitaries and carrying out a quantum measurement detecting the spin projections along $x$, $y$, or $z$. A total of $N_U=498$ different unitaries was used; for each unitary, the quantum state was $N_M=150$ times prepared and measured.   

\begin{figure}
    \centering
    \includegraphics[width=.5\columnwidth]{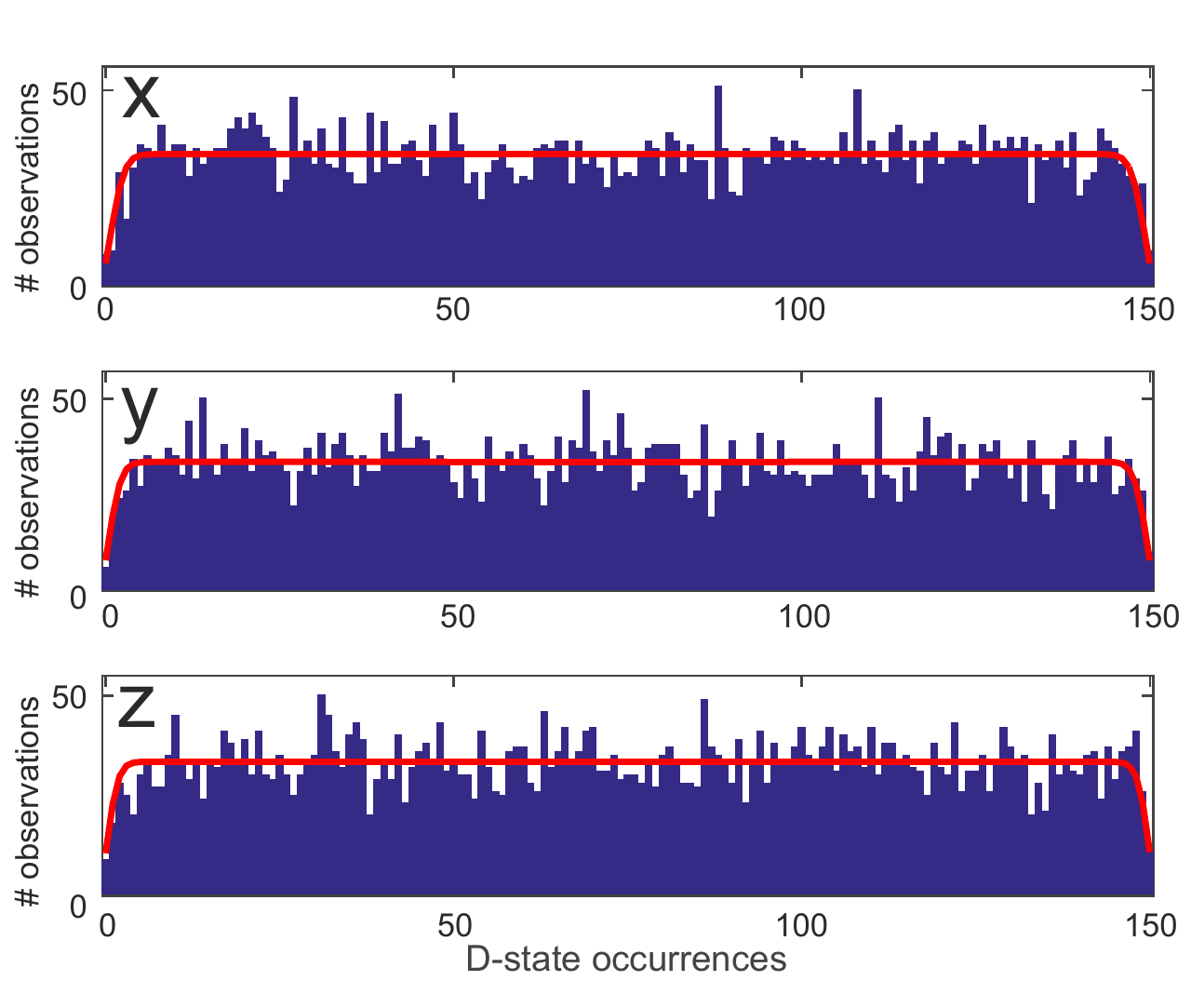}
    \caption{\small{\textbf{Single-qubit spin projection distributions onto X,Y,Z after applying a random spin rotation to a qubit in a pure state.}} Measured histograms (blue bars) are shown together with a fit (red line) taking account decoherence and the finite number of measurements.}
    \label{fig:probability-distributions-XYZ}
\end{figure}

The resulting spin projection distributions are shown in Fig.~\ref{fig:probability-distributions-XYZ} where the histograms show the number of occurences of finding an ion $m$ times ($0\le m \le N_M$) in the non-fluorescing D-state. As can be seen, the histograms are reasonably flat, however the distribution falls off towards the extreme values, indicating that the state is not perfectly pure.

We model the decoherence as a depolarizing channel, $\rho\rightarrow \lambda\rho + \frac{1-\lambda}{2}{\cal I}$,
and fit the probability distribution $f(p)$ of finding the ion in the D-state with a box-like distribution ($f(p)=0$ for $p<p_{lim}$ and $p>1-p_{lim}$, $f(p)=1/(1-2p_{lim})$ for $p_{lim}<p<1-p_{lim}$) that is convoluted with quantum projection noise. We find $p_{lim}^x=0.0115$, $p_{lim}^y= 0.0095$, $p_{lim}^z=0.0075$ when detecting in X, Y, or Z. This corresponds to an average loss of purity of $\gamma=0.019$ per qubit, resulting in a reconstructed purity of $(1-\gamma)^{10}\approx 0.83$ for a 10-qubit product state.
For testing the goodness-of-fit, we carried out a $\chi^2$-test yielding $\chi^2_X=142$, $\chi^2_Y=160$, $\chi^2_Z=153$, all of which were consistent with the expected value of 149.

\begin{figure}
    \centering
    \includegraphics[width=1\columnwidth]{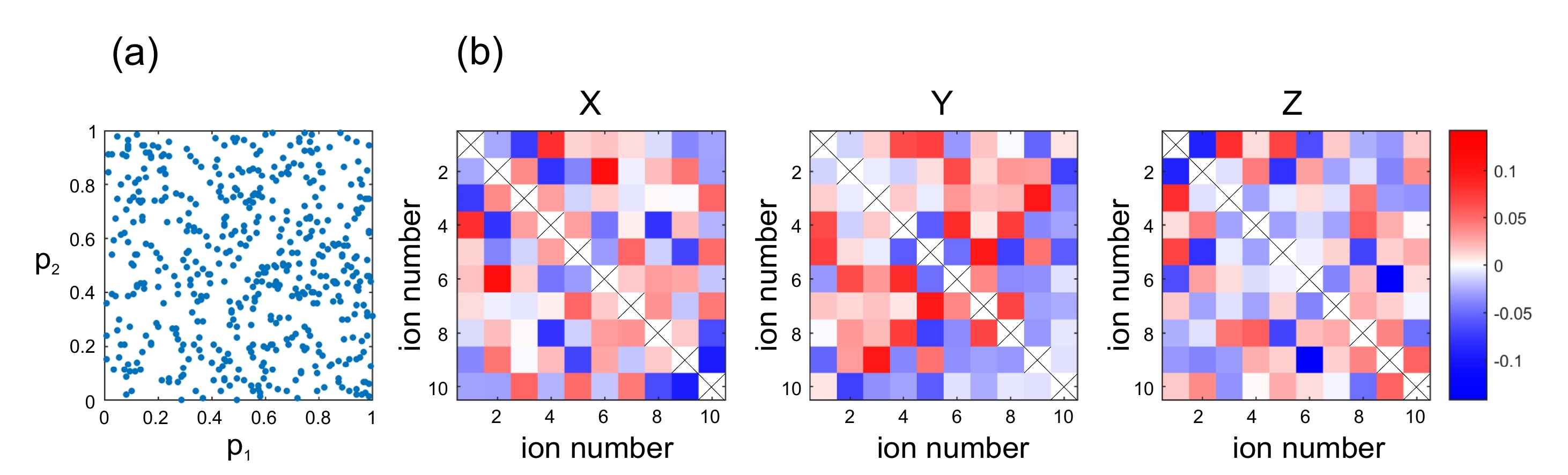}
    \caption{\small{\textbf{Single-qubit randomized measurement results on a 10-ion string}}. In this measurement, all ions were prepared in the electronic ground state. (a) Probability $p_1$ for ion 1 being found in the excited state versus probability $p_2$ for ion 2 being excited. (b) Correlation coefficients for probabilities ($p_i$, $p_j$) when measuring in the $X$, $Y$, and $Z$ bases.}
    \label{fig:CorrelationsBetweenDifferent Ions}
\end{figure}

We also tested whether cross-talk between neighbouring ions induced by imperfect focusing of the strongly focused laser beam could give rise to correlations between the recorded probabilities on different ions. Fig.~\ref{fig:CorrelationsBetweenDifferent Ions} (a) shows an example of probability pairs ($p_1^z$, $p_2^z$) of finding ion 1 and ion 2 in the excited state for 498 different local random unitaries (we use the same data set as in the previous figure) for a measurement in the z-basis. No obvious correlations are discernible in this example. To better quantify potential correlations, we calculated the Pearson correlation coefficient $c_{ij}^\alpha=\mbox{cov}(p_i^\alpha,p_j^\alpha)/(\sigma_{p_i^\alpha}\sigma_{p_j^\alpha})$ between the probabilities $p_i^\alpha$ ($p_j^\alpha$) of ion $i$ ($j$) being found in the excited state, as shown in Fig.~\ref{fig:CorrelationsBetweenDifferent Ions} (b) for measurements in the $x$-,$y$-, and $z$-basis ($\alpha\in\{x,y,z\})$.
The non-zero correlation coefficients are not significant. Moreover, we applied Fisher's combined probability test to the p-values of the correlation coefficients $c_{ij}$, by calculating
$ \chi^2=-2\sum_{i<j}\log p_{ij}$.
For a Fisher test comprising $M$ different p-values, a value of $\chi_{\mathrm{uncorr}}^2=2M$ would be expected for uncorrelated probabilities. A numerical simulation assuming such uncorrelated probabilities yielded $\chi_{\mathrm{sim}}^2=270\,(23)$. For our experiments, we found $\chi^2=261$, from which we conclude that there were no significant correlations between the operations carried out on different ions that could be detected by this test.

\subsection{Many-body quantum dynamics}

\textbf{Magnetization dynamics under the evolution of $H_{\mathrm{XY}}$, with and without disorder}

\begin{figure}
    \centering
    \includegraphics[width=.7\columnwidth]{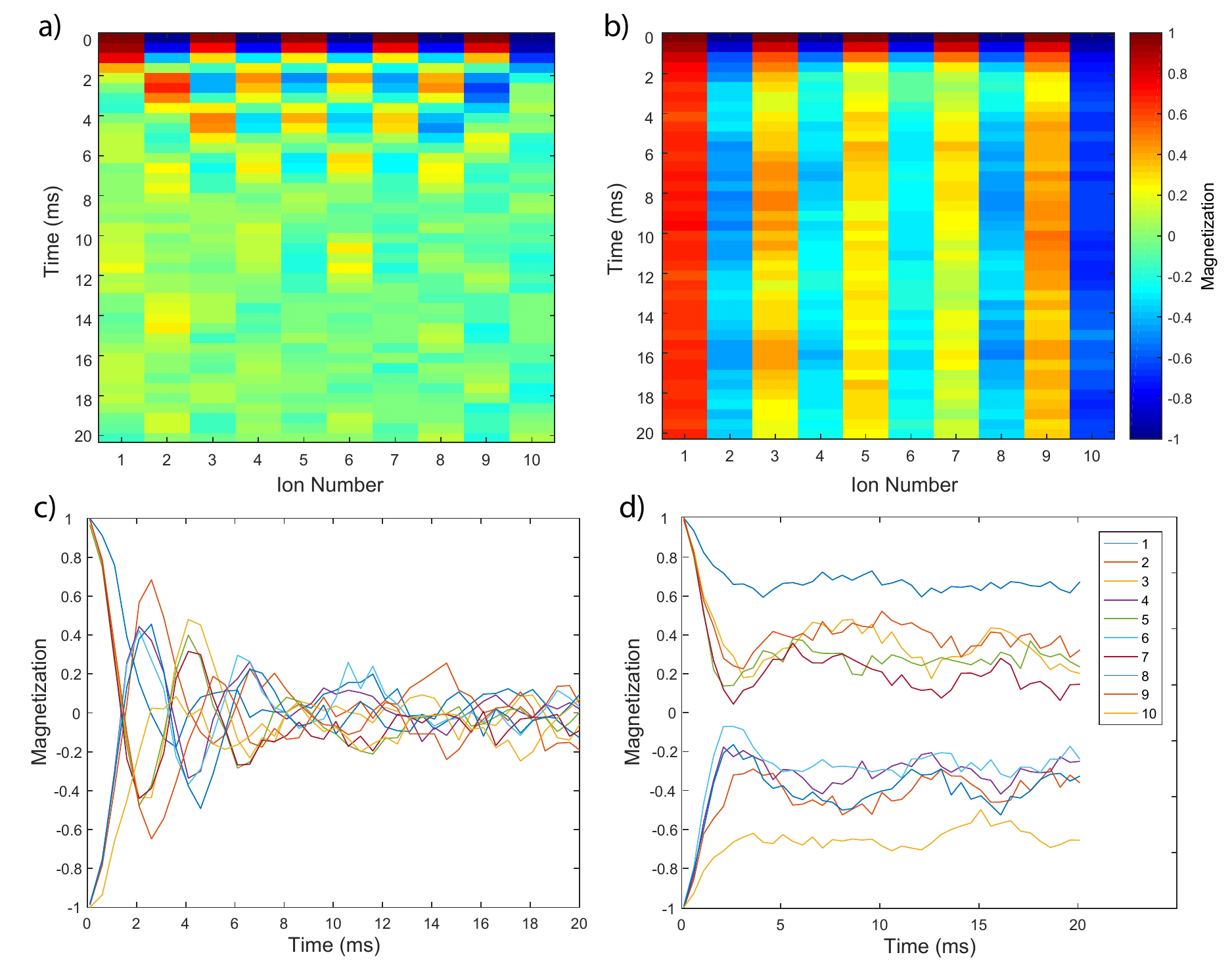}
    \caption{\small{\textbf{Magnetization evolution.} (a) Time evolution of the 10 ion initial N\'eel ordered state under the Hamiltonian $H_{\mathrm{XY}}$ (no disorder). (b) Time evolution of the 10 ion initial N\'eel ordered state with on-site disorder. Averaging is performed over the 35 random realizations implemented in the experiment (see main text). (c) and (d) Spatially resolved z-magnetization for 10 ions with no disorder (left) and with disorder (right).}}
    \label{fig:walks}
\end{figure}

The dynamical evolution of the magnetization $\langle Z_{i}\rangle$ (proportional to the probability of finding a qubit excitation at site $i$) shows how the multiple excitations of the initial 10 ion N\'eel ordered state disperse under the application of $H_{\mathrm{XY}}$. Figures~\ref{fig:walks} (a) and (b) show this evolution for no disorder and with disorder present. The magnetization dynamics were obtained by applying an instance of disorder and time evolving the state. This was repeated for all 35 randomly drawn disorder patterns, and the dynamics averaged over the disorder (Fig.~\ref{fig:walks} (b)). The averaged dynamics retain many of the same characteristics of the initial N\'eel ordered state, indicating that there is a remembrance of the initial state during the dynamics. 

Figures~\ref{fig:walks} (c) and (d) show the corresponding evolution of the single-spin magnetization, both without disorder and in the presence of disorder. With no disorder present, the initially localized excitations rapidly disperse throughout the system, resulting in an approximately equal magnetization for all ions at longer times. In the presence of disorder, a stationary magnetization is observed, showing evidence of a localized phase.\\

\begin{figure}
    \centering
    \includegraphics[width=.6\columnwidth]{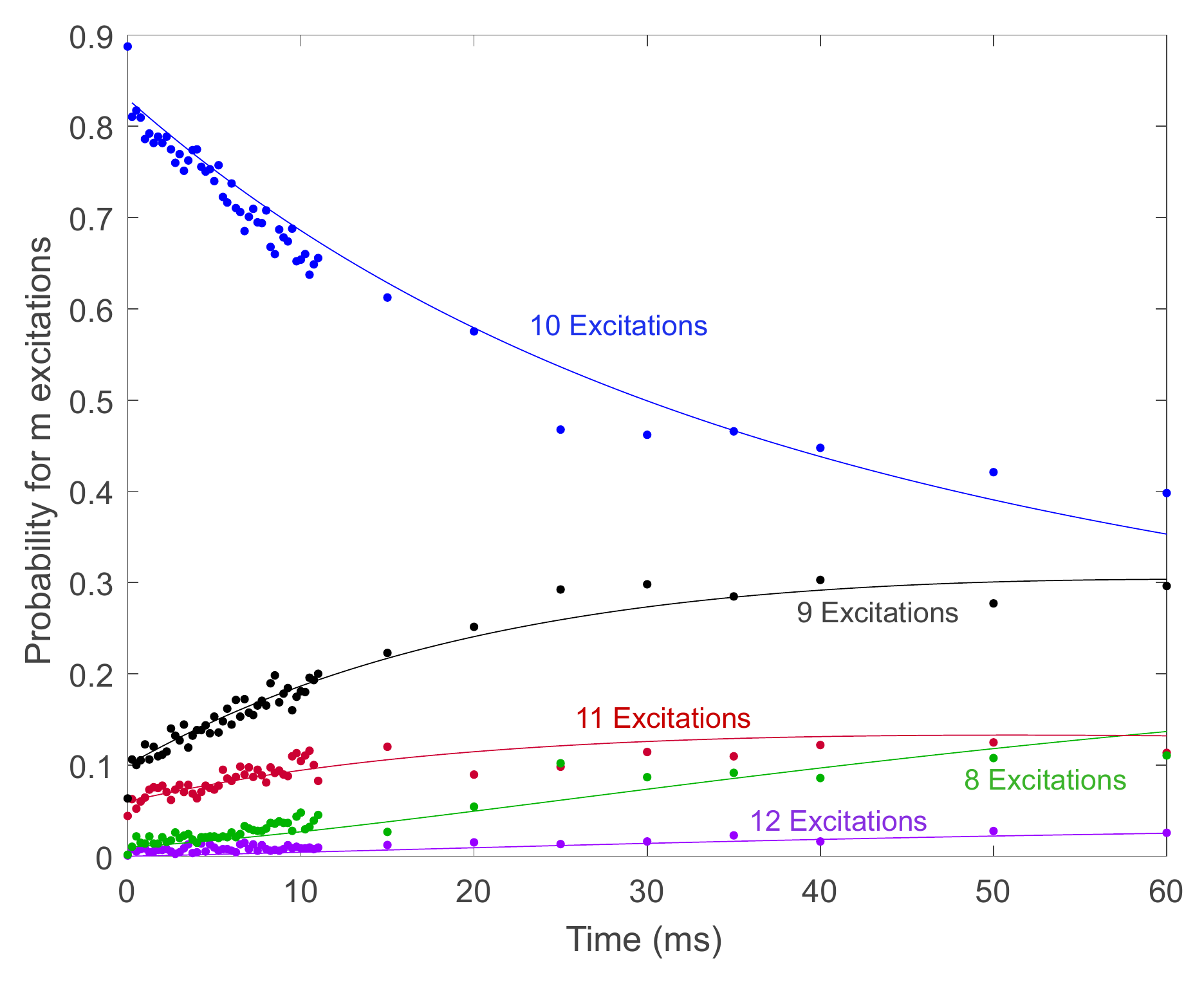}
    \caption{\small{\textbf{Excitation number dynamics under the XY Hamiltonian, starting from the N\'eel ordered state.} A fit of the model (lines) to the data (points) is shown for 8 to 12 excitations.}}
    \label{fig:exc_dyn}
\end{figure}

\noindent\textbf{Excitation number dynamics as an indicator of decoherence}\\
In a time evolution governed by an ideal XY Hamiltonian, the number of excitations should be conserved. However, experimentally it is observed that the excitation number is not strictly conserved, as shown by the data points in Figure~\ref{fig:exc_dyn}. Here, a 20-ion N\'eel ordered state is evolved under $H_{\mathrm{XY}}$ from 0 to 60~ms. This lack of conservation can occur for two main reasons: 1. The finite lifetime of the D$_{5/2}$ state results in decay to the S$_{1/2}$ state. 2. Imperfections in the laser-ion interaction, for example high-frequency laser phase noise or a disruption of the spin-spin coupling due to motional heating, give rise to spin flips. The excitation number dynamics can be modelled by assuming a spontaneous decay rate of $\Gamma$, and an additional incoherent spin flip rate, $\gamma_{flip}$, which is independent of the electronic state. As such, the probability, $p$ for one ion to be in the excited state evolves according to:
\begin{equation}\label{excited_evolution} 
\dot{p}=-(\Gamma + \gamma_{flip})p + \gamma_{flip}(1-p)
\end{equation} 
This equation has a solution of the form $p(t) = p_{eq}+(p_{i}-p_{eq})e^{-\lambda t}$, where $\lambda=2\gamma_{flip}+\Gamma$, $p_{eq}=\gamma_{flip}/\lambda$ is the steady-state probability, and $p_{i}$ is the probability of being initially in the excited state.\\
This model can be extended to accommodate the dynamics with $N$ ions, assuming that initially $N_{1}$ ions are in the excited state, each with probability $p_{1}(t)=p_{eq}+(1-p_{eq})e^{-\lambda t}$ to be found in the excited state. Consequently there will be $N_{2}=N-N_{1}$ ions in the electronic ground state, each with probability $p_{2}(t)=p_{eq}(1-e^{-\lambda t})$ to be in the excited state.\\
The probability for $k$ ions to be excited at a given time $t$, $p_{k}(t)$, can be expressed as $k = k_{1} + k_{2}$. Here, $k_{1}$ represents the ions which were initially excited, and $k_{2}$ the ions which were initially in the ground state. $p_{k}(t)$ can then be expressed in terms of $p_{1}$ and $p_{2}$ such that:
\begin{equation}\label{exc_dyn_sim}
    p_{k}(t) = \sum^{k_{max}}_{k_{1}=k_{min}}p_{1}^{k_{1}}(1-p_{1})^{N_{1}-k_{1}}p_{2}^{k_{2}}(1-p_{2})^{N_{2}-k_{2}}\begin{pmatrix}N_{1}\\ k_{1}\end{pmatrix}\begin{pmatrix}N_{2}\\ k_{2}\end{pmatrix}
\end{equation}
where $k_{min} = max(0,k-N_{2})$ and $k_{max} = min(k,N_{1})$. Fig.~\ref{fig:exc_dyn} fits Eq. (\ref{exc_dyn_sim}) to the data (lines) assuming a spontaneous decay rate of $\Gamma =\tau^{-1}$ with $\tau=1.17$~s. This fit is optimal for a single-ion spin flip rate of $\gamma_{flip}=0.69\,\mathrm{s}^{-1}$. As such, for a 20-ion Coulomb crystal prepared in the N\'eel-ordered state and illuminated by a bichromatic beam with 40~kHz detuning, it would take on average 70\,ms for an unwanted spin flip caused by laser-ion interactions to occur, whereas spin flips by spontaneous decay occur about every 100~ms.\\

\begin{figure}
    \centering
    \includegraphics[width=0.5\columnwidth]{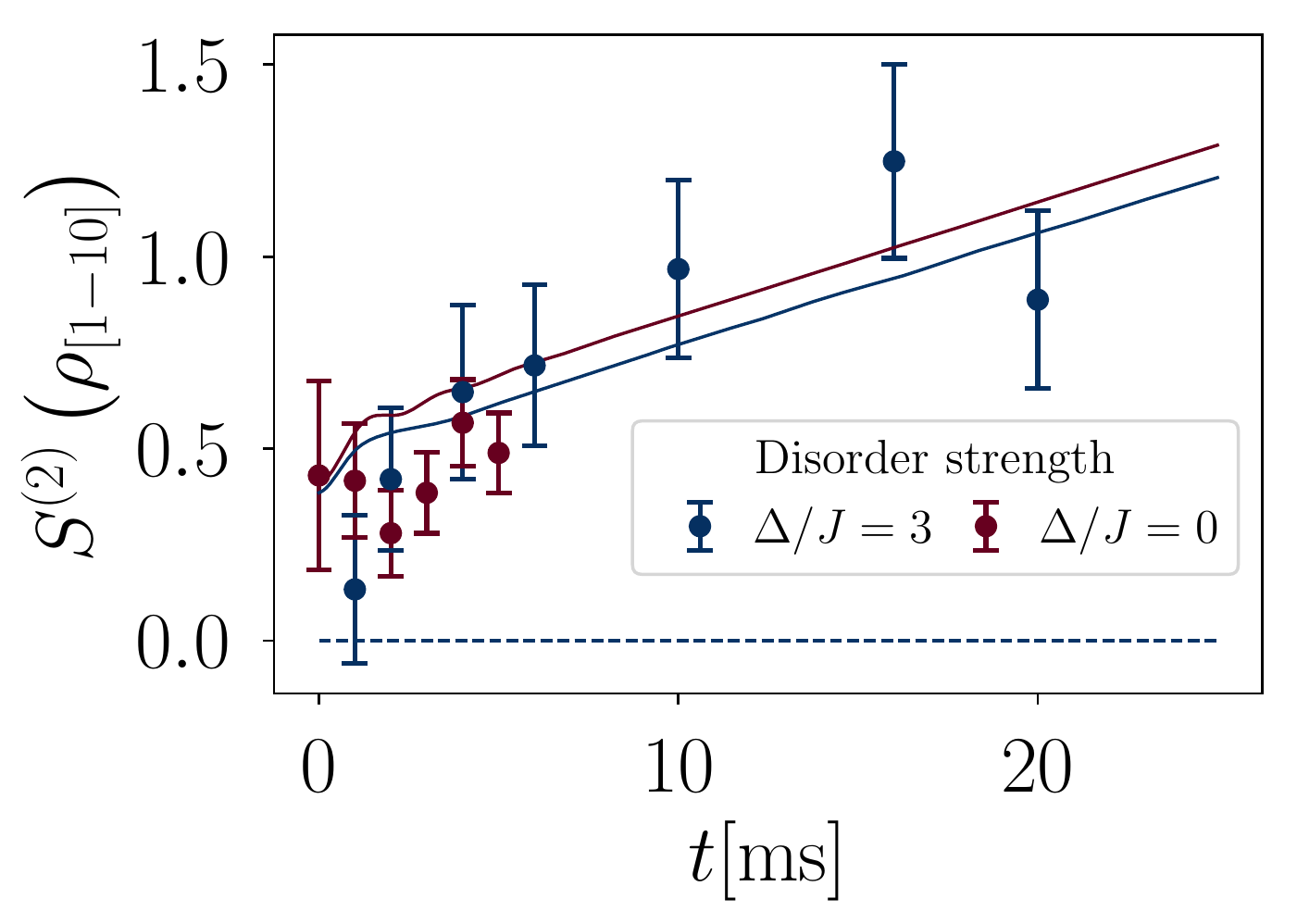}
    \caption{{\bf Evolution of the full system entropy under $H_{\mathrm{XY}}$ ($J_0=420\,$s$^{-1}$, $\alpha=1.24$) with and without disorder.} Full system entropy $S^{(2)}(\rho_{[1\rightarrow 10]})$ vs.\ time without disorder (red data points) and with disorder (blue data points). The parameters are the same as for Fig.\ 4 of the main text. Numerical simulations based on unitary dynamics (dotted curves) and including known sources of decoherence (full lines) are in agreement with the measured R\'enyi entropies. Both dotted lines lie at zero for the duration of the dynamics.}
    \label{fig:full_system_entropy}
\end{figure}

\noindent\textbf{Dynamics of the full 10-qubit system entropy with and without disorder}\\
Fig.~\ref{fig:full_system_entropy} shows the temporal evolution of the full 10-qubit system entropy under the XY-Hamiltonian without disorder (the same data is also plotted in the 10-qubit partition column of Fig.~\ref{fig:10-qubit Purity-and-entropy vs partition size}(b)) as well as the corresponding entropy evolution for the case of disorder. As can be seen, the measured entropy remains basically constant in the case without disorder over the measured time span. In the case with disorder, there is a slight entropy increase with time up to entropy values that are, however, small compared to the half-chain entropy (see Fig.~\ref{fig:data-with-disorder}(a) of the main text.). 

\subsection{Numerical simulations}
In this section, we give additional details about the numerical results presented in the manuscript.

For the simulations with a total number of $10$ qubits, we solve the master equation exactly~\cite{Johansson:2013}, using the $J_{ij}$ matrix, calculated using the parameters of the experiment, as in Ref.\ \cite{Jurcevic:2014}, and including all sources of decoherence mentioned above: imperfect preparation of the N\'eel state, spontaneous emission and spin flips during time evolution, and depolarization due to the local unitary operations.

For the 20-qubit simulation, we use a 
matrix-product state (MPS) algorithm which represents the time-evolved quantum state in a compressed, factorized form, whose size is controlled by the choice of the bond dimension $D$. The simulations were performed using the ITensor library (http://itensor.org). To treat the effect of decoherence, we use quantum trajectories~\cite{Daley:2014}, which corresponds to adding to the Hamiltonian a local non-Hermitian component, and to subjecting the MPS during time evolution to random local quantum jumps.
Regarding the unitary part of the evolution, our code is based on representing the $J_{ij}$ matrix as a sum of $n_e$ exponentially decaying terms, which can be represented efficiently in the Matrix-Product-Operator (MPO) language~\cite{Pirvu:2010}. Time evolution is then implemented using a first-order Trotter approximation~\cite{Zaletel:2015}.
For the simulations presented in Fig.~3, we considered $n_e=3$ exponential terms to represent the $J_{ij}$ matrix, and used $100$ quantum trajectories, each obtained with maximum bond dimension $D=192$.

\end{document}